\documentclass[published]{aa}
\usepackage{booktabs}
\usepackage{graphicx}
\usepackage{txfonts}
\usepackage[hidelinks]{hyperref}
\usepackage{xcolor}

\newcommand{\Eq}[1]{Eq.~(\ref{#1})}
\newcommand{\Eqs}[2]{Eqs.~(\ref{#1}) and (\ref{#2})}
\newcommand{\Fig}[1]{Fig.~\ref{#1}}

\begin{document}

\title{A model of umbral oscillations inherited from subphotospheric fast-body modes}

\author{Juhyung Kang\inst{1}
    \and Jongchul Chae\inst{1}
    \and Kyuhyoun Cho\inst{2,3}
    \and Soosang Kang\inst{1}
    \and Eun-Kyung Lim \inst{4}
}
\institute{Astronomy Program, Department of Physics and Astronomy, Seoul National University, Seoul 08826, Republic of Korea\\
    \email{jcchae@snu.ac.kr}
    \and Bay Area Environmental Research Institute, NASA Research Park, Moffett Field, CA 94035, USA
    \and Lockheed Martin Solar and Astrophysics Laboratory, 3251 Hanover St, Palo Alto, CA 94306, USA
    \and Solar and Space Weather Group, Korea Astronomy and Space Science Institute, Daejeon 34055, Republic of Korea}

\date{Received MM DD, YYYY /  Accepted MM DD, YYYY}

\abstract{
    Recently, complex horizontal patterns of umbral oscillations have been reported, but their physical nature and origin are still not fully understood.
    Here we show that the two-dimensional patterns of umbral oscillations of slow waves are inherited from the subphotospheric fast-body modes.
    Using a simple analytic model, we successfully reproduced the temporal evolution of oscillation patterns with a finite number of fast-body modes.
    In this model, the radial apparent propagation of the pattern is associated with the appropriate combination of the amplitudes in radial modes.
    We also find that the oscillation patterns are dependent on the oscillation period.
    This result indicates that there is a cutoff radial mode, which is a unique characteristic of the model of fast-body modes.
    In principle, both internal and external sources can excite these fast-body modes and produce horizontal patterns of umbral oscillations. 
    }


\titlerunning{Subphotospheric Fast Body Modes}
\authorrunning{Kang et al.}

\maketitle

\section{Introduction} \label{sec:intro}
Umbral oscillations are a magnetohydrodynamic (MHD) process conspicuous in every sunspot umbra.
Following their initial detection \citep{Beckers1969}, subsequent works revealed that the umbral oscillations are the slow MHD waves propagating upwards from the photosphere to the corona with a group speed of around the speed of sound \citep{Lites1984,Centeno2006,Felipe2010,KP2017}.
Because the temperature minimum region between the photosphere and the chromosphere acts as a high-pass filter, the wave power peaks around the cutoff frequency ($\omega_c$) of around 6 mHz in the chromosphere \citep{Roberts1983,Centeno2006}.
Theoretical studies have revealed that even a portion of low-frequency waves ($\omega < \omega_c$) can propagate upwards due to the effects of the temperature gradient \citep{Schmitz1998,Chae2018} and the radiative relaxation of nonadiabatic heating and cooling \citep{Roberts1983,Centeno2006,Chae2023}.
The most plausible sources of the umbral oscillations have been considered to be either the magnetoconvection inside a sunspot, such as an umbral dot \citep{Lee1993, Jacoutot2008, Jess2012}, or the absorption of the incident $f$ and $p$ mode waves \citep{Braun1987,Spruit1992,Cally1997}.

Interestingly, observational studies recently reported waves that appear to move across the magnetic field, forming ring-like patterns \citep{Zhao2015,Cho2020,Cho2021} or spiral-shaped wave patterns (SWPs) \citep{Sych2014,Su2016,Felipe2019,Kang2019}.
There are two models that can reproduce these patterns: a model of propagating waves excited in the subphotosphere \citep[e.g.,][]{Zhao2015,Cho2020}, and a model of slow-body resonant modes \citep[e.g.,][]{Edwin1982, Stangalini2022}.
The wave-propagation model assumes that there is a localized disturbance in the high-$\beta$ region of the subphotosphere. This disturbance excites fast MHD waves, and these waves propagate quasi-isotropically.
The wavefront reaching the equipartition layer ($\beta \sim 1$) has a time delay as a function of horizontal distance, and this time delay causes ring-like patterns to propagate across the magnetic field \citep{Zhao2015}.
After reaching $\beta \sim 1$, the fast waves are partially converted to the slow waves by the mode-conversion process \citep{Zhugzhda1984,Cally2001,Schunker2006}.
This model successfully reproduces the ring-like wavefronts \citep{Cho2020,Cho2021} and the SWPs \citep{Kang2019} observed in the chromosphere.
To explain the SWPs, \citet{Kang2019} additionally assumed that the point-like source generates nonaxisymmetric modes.
However, it is not yet clear as to whether or not the exciting source assumed in this model really exists, and if it does, how such a source can generate nonaxisymmetric modes also remains unclear.

The other model to explain the complex wave patterns is the model of slow-body resonant modes.
Eigenmode
waves are known to appear when the flux tube resonates with external driving  \citep{Edwin1982,Edwin1983,Roberts2019}.
Considering the cylindrical geometry, \citet{Jess2017} successfully identified the first-order azimuthal mode from chromospheric umbral oscillations.
Recently, \citet{Albidah2022} found higher-mode oscillation patterns in circular and elliptical sunspots.
In addition, in a large-scale sunspot with complex geometry, \citet{Stangalini2022} successfully reproduced the oscillation patterns by the superposition of more than 30 slow-body resonant modes in Cartesian geometry.
However, it is difficult to understand how a systematic horizontal pattern can be established from the resonance of slow-body waves that are very inefficient in transferring energy in the horizontal direction across magnetic fields.

In the present work, we propose another model: a model of umbral oscillations inherited from subphotospheric fast-body modes.
In this model of fast-body modes, we suppose that the driving and resonance of the flux tube occur in the subphotosphere, resulting in the subphotospheric fast-body modes.
These fast modes are partially converted to slow waves in the equipartition layer.
These slow waves come to have two-dimensional patterns with resonant modes because they inherit the fast modes from the subphotosphere.

In Sect. \ref{sec:model}, we propose the simple analytic model of the subphotospheric fast-body modes.
In Sect. \ref{sec:method}, we show observations of two-dimensional patterns in the pore and present a method to analyze them.
In Sect. \ref{sec:res}, we compare the observation and the model.
Finally, in Sect. \ref{sec:dis}, we summarize and discuss our model.

\section{The model} \label{sec:model}
\begin{figure*}
    \includegraphics[width=\hsize]{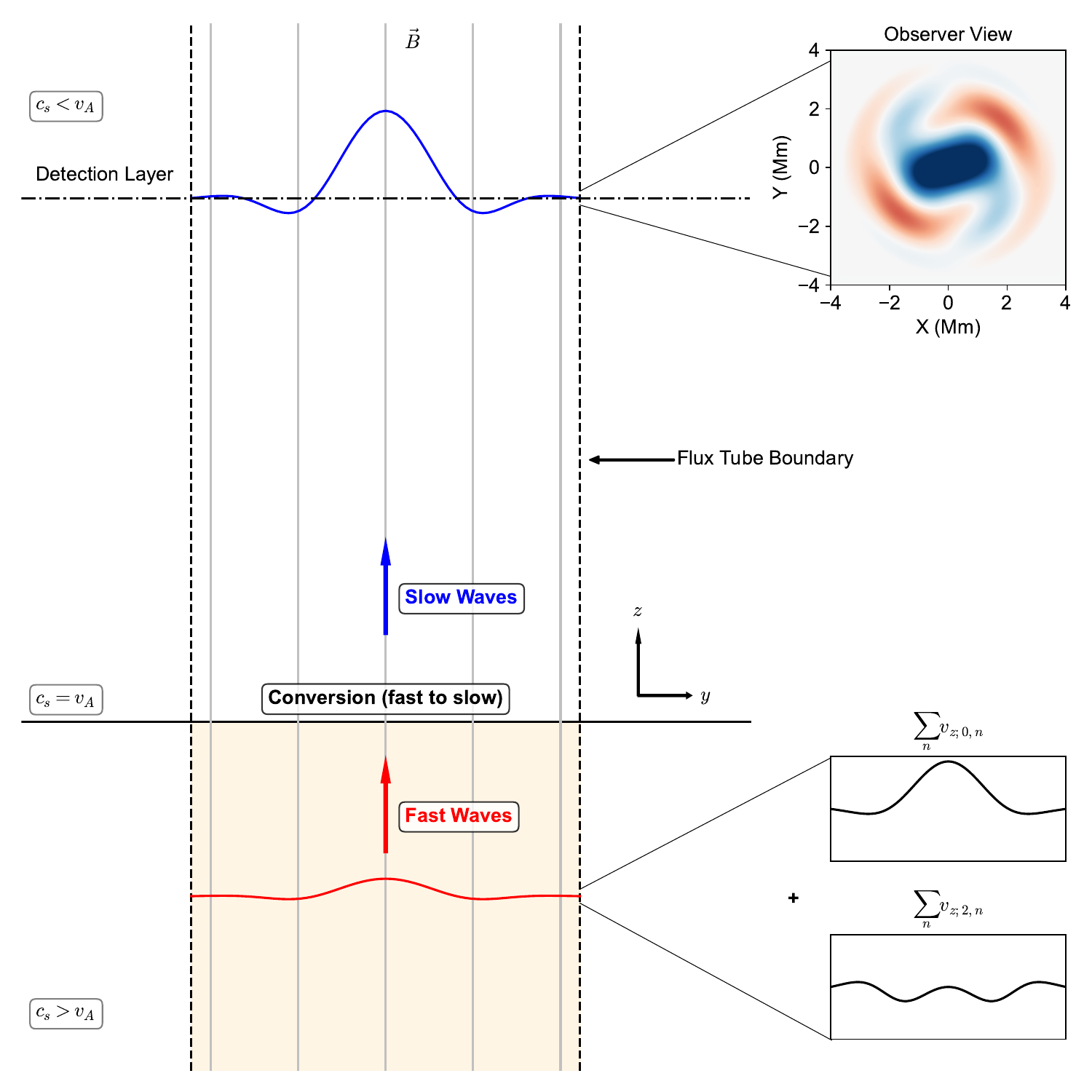}
    \caption{Illustration of our model of umbral oscillations inherited from subphotospheric fast-body modes.
             The yellow area represents the high-$\beta$ region ($c_s > v_A$) below the sunspot surface, and the white area represents the low-$\beta$ region ($c_s < v_A$).
             The black solid line between the two regions represents the equipartition layer ($c_s = v_A$) where the mode conversion can occur.
             The gray vertical lines illustrate the magnetic field lines of the flux tube, and the vertical dashed line indicates the boundary of the flux tube.
             The red line demonstrates the vertical velocity $v_z$ of fast body waves, and the blue line represents $v_z$ of {converted} slow waves.
             The dash-dotted horizontal line represents the detection layer of the chromosphere.
             The two panels in the bottom right corner show the vertical velocity fluctuation of two azimuthal modes of $m=0$ and 2.
             The top right panel shows a horizontal cross-section view of the modeled oscillation patterns.
             }\label{fig:schem}
\end{figure*}

We regard the patterns of umbral oscillations as the inherited patterns of fast-body modes in the subphotosphere (see Figure \ref{fig:schem}).
We conjecture that the inheritance is realized by the process of fast-to-slow-mode conversion in the equipartition layer \citep[e.g.,][]{Zhugzhda1984,Cally2001} after the fast-body modes have formed in the subsurface.
These converted slow waves (blue solid line) propagate upwards to the detection layer along the magnetic field.
As a consequence, the observed horizontal patterns of umbral oscillations can be identified with the patterns of fast-body modes (red solid line) in the subphotosphere.

For simplicity, we adopt a vertical magnetic flux tube with a radius $R$ where density $\rho$, Alfv\'en speed $v_A$, sound speed $c_s$, and tube speed $c_T \equiv c_s v_A/\sqrt{c_s^2 + v_A^2}$ are uniform inside the flux tube.
The magnetoacoustic body mode of vertical velocity with angular frequency $\omega$, azimuthal mode $m$, radial wavenumber $k_r$, and vertical wavenumber $k_z$ is given by the solution in cylindrical coordinates \citep{Edwin1983,Roberts2019},
\begin{equation}
        v_z  (t, r, \theta, z; \omega,  k_r, m, k_z) = A_{k_r, m, \omega} J_m(k_r r) e^{i\left(k_zz + m\theta - \omega t\right)},  ~~~0 \leq r < R \label{eq:vz}
,\end{equation}
where $A_{k_r, m, \omega}$ is a complex constant and $J_m$ is the first kind of the Bessel function of an azimuthal mode $m$.
Here, the radial wavenumber $k_r$  depends on $\omega$ and $k_z$ as in the expression
\begin{equation}
        k_r^2 = - \frac{(k_z^2 c_s^2 - \omega^2)(k_z^2 v_A^2 - \omega^2)}{(c_s^2 + v_A^2)(k_z^2 c_T^2 - \omega^2)} \label{eq:kr}
,\end{equation}
and is to be determined from the boundary condition at $r=R:$
\begin{equation}
    \frac{1}{\rho_e (k_z^2 v_{A,e}^2 - \omega^2)} k_{r,e} \frac{K'_m(k_{r,e} R)}{K_m(k_{r,e} R)} = \frac{1}{\rho (k_z^2 v_A^2 - \omega^2)} k_r \frac{J'_m(k_r R)}{J_m(k_r R)}  \label{eq:dispersion}
,\end{equation}
which is obtained from the continuity of total pressure and radial velocity $v_r$ \citep{Edwin1983}.
Here, $K_m$ is the second kind of the modified Bessel function, the prime operator ($\prime$) denotes the derivative of the Bessel function, and the subscript $e$ stands for quantities in the external region $r>R$.

For the body wave solution, $k_r$ should be real and hence $k_r^2$ has to be positive \citep{Edwin1983}.
This requirement is satisfied for fast-body waves when the vertical phase speed $v_p \equiv \omega/k_z$ lies between the internal sound speed and the external sound speed:
\begin{equation} \label{eq:fastcon}
c_{s,e} > v_p > c_s > v_A \, . 
\end{equation}
Similarly, we obtain the condition for the slow-body wave solution:
\begin{equation}
v_A > v_p > c_T
,\end{equation}
indicating that $v_p$ should be in between the internal Alfv\'en speed and the tube speed.

We note that Equation~\ref{eq:dispersion} has a simple solution,
\begin{equation}
J_m ({k_r(n)} R) = 0 \label{eq:simple}
,\end{equation}
in the limit of zero-external density ($\rho_e=0$).
This condition might be satisfied when the external medium is much hotter than the interior.
As $J_m$ is an oscillatory function, there are several values of {$k_r(n)$} with different values of radial {mode} $n$.
{In other words, $k_r(n)R$ is the $n$-th root of \Eq{eq:simple}.}
Generally speaking, $k_r$ is specified by $m$, $n,$ {and $R$}.

The zero-density limit is useful for understanding the characteristics of the body waves when only considering the internal atmospheric conditions such as $c_s,$ and $v_A$.
\Eqs{eq:kr}{eq:simple} are solved for the expression
\begin{equation}
    k_{z,\pm}^2 = \frac{-b \pm \sqrt{b^2 - 4 a c}}{2a}, \label{eq:kz}
\end{equation}
with the parameters
\begin{eqnarray}
    &a &= v_A^2 c_s^2, \label{eq:a} \\
    &b &= (c_s^2 + v_A^2)(k_r^2(n) c_T^2 - \omega^2), \label{eq:b} \\
    &c &= \omega^4 - k_r^2(n) \omega^2 (c_s^2 +v_A^2). \label{eq:c}
\end{eqnarray}
We note that the larger value, $k_{z,+}$ , corresponds to slow waves and the smaller one, $k_{z,-}$, to fast waves.

\begin{figure*}
    \includegraphics[width=\hsize]{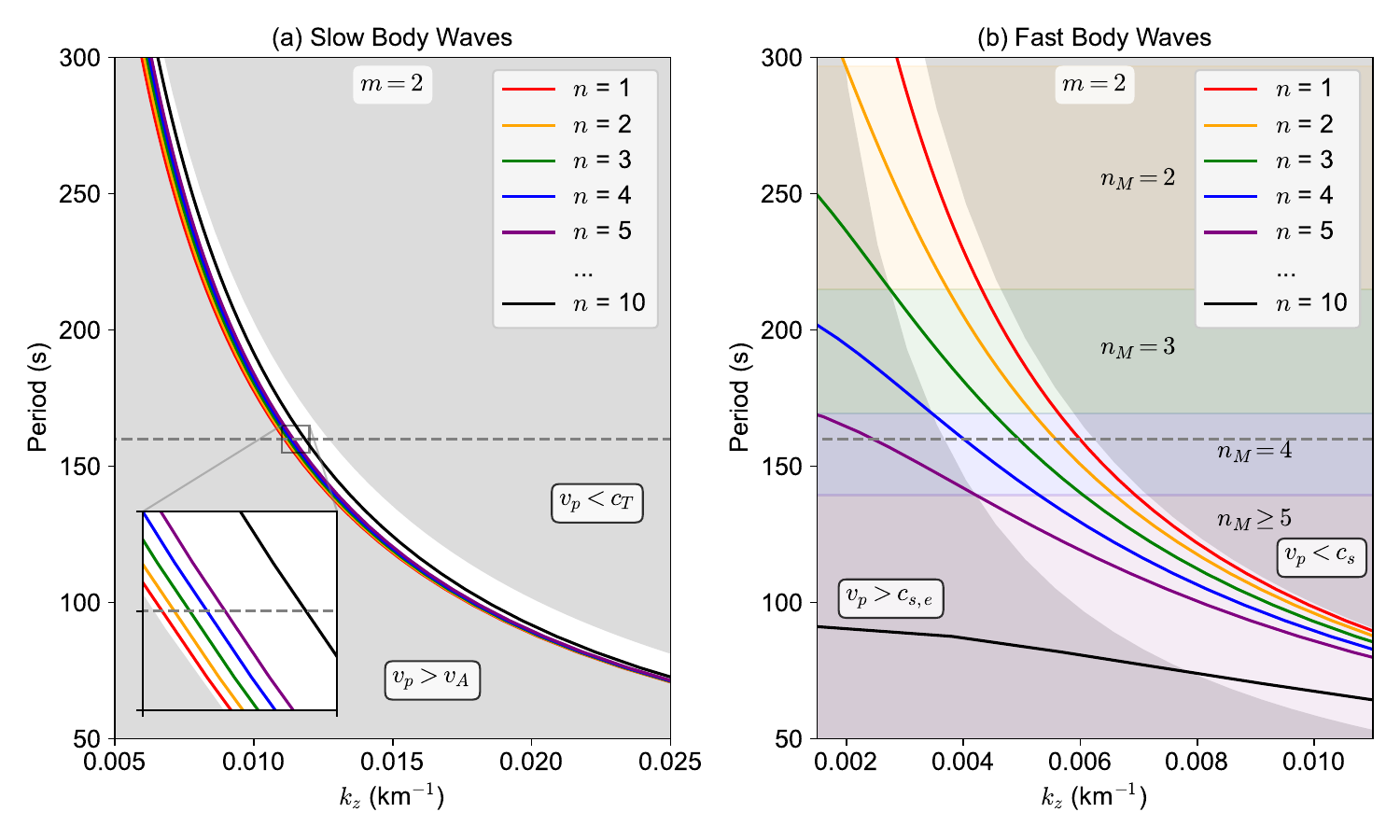}
    \caption{Period $P$ as a function of vertical wavenumber $k_z$ for (a) the slow-body waves and (b) the fast-body waves for $n=1, 2, 3, 4, 5$, and $10$ with $m=2$ in the subphotospheric condition ($c_{s,e} > c_s > v_A$).
    The white area in panel (a) marks the range of $v_A > v_p > c_T$ within which the slow-body wave solution can exist, and that in panel (b) represents the range within
which the fast-body wave can exist ($c_{s,e} > v_p > c_s$).
    The inset figure in panel (a) shows the magnified figure in the ranges of $0.011<k_z<0.012$ km s$^{-1}$ and $155<P<165$ s.
    The dashed line represents the period of 160 seconds.
    {The color-shaded regions in panel (b) represent the regions with the same cutoff radial mode $n_M$; yellow is $n_M=2$, green is $n_M=3$, blue is $n_M=4$ and purple is $n_M\ge5$.}
    We have taken the internal sound speed of $c_s = 6.3$ km s$^{-1}$, internal Alfv\'en speed of $v_A = 0.56 c_s$, and external sound speed of $c_{s,e} = 1.7c_s$.
    }\label{fig:pk}
\end{figure*}

The relationship in Equation \ref{eq:kz} can be used to calculate $k_z$ as a function of $\omega$ or equivalently the oscillation period $P=2\pi/\omega$ when $m$ and $n$ are specified in the cases of slow-body waves and fast-body waves.
As an illustration, we consider a flux tube of $R = 3.5$ Mm, $c_s = 6.3 $ km s$^{-1}$, and $v_A=0.56c_s$ at a depth of $\beta = 3.8$ below the surface (Figure \ref{fig:pk}).
Here, the choice of $c_s$ has been made by extrapolating the Maltby M sunspot model \citep{Maltby1986} to the depth of 100 km.
As the condition $c_{s,e} > c_s > v_A$ is satisfied in this flux tube, both fast-body waves and slow-body waves can occur if $v_p$ is within a specified range.

We first consider slow-body modes (see Figure \ref{fig:pk}a).
These waves can exist within the $v_p$ range marked as a white area and satisfying $v_A > v_p > c_T$.
We find that the curve representing the solution for $k_z(P)$ with $m=2$ of any value of $n$ fits into this range.
An important property of slow waves is that for a specified period, there exists an infinite number of solutions with $n=1, ..., \infty$ with any value of $m$.

We now consider fast-body modes.
Figure \ref{fig:pk}b presents the solution for $k_z(P)$ for each value of $n$ with $m=2$.
Even though this solution is independent of the physical condition outside the flux tube, it can have a physical meaning as fast waves only when the condition $c_{s,e} > v_p > c_s $ is satisfied.
With the choice of $c_{s,e} = 1.7 c_s$, the physically meaningful range of $v_p$ is marked as a white area.
A distinct property of fast waves is that for a specified period, there exists a finite number of solutions with $n \leq n_M$ for each $m$, where $n_M$ is {the highest radial mode that can be trapped in the flux tube satisfying the condition of \Eq{eq:fastcon}. }
In other words, for a  given period, there exists a cutoff {($n_M$)} in the radial mode in fast-body waves.
{For example, for the waves with a period of 160 s, which corresponds to the dashed line, the cutoff radial mode is $n_M=4$.}
We note that $n_M$ depends on $P$ as well as $m$ in fast waves, whereas $n_M = \infty$ always in slow waves.

As we regard the observed pattern of velocity fluctuation as the inheritance of the fast-body waves in the subphotosphere, it can be modeled by a superposition of all modes, as in
\begin{eqnarray}
    \lefteqn{v_z (t, r, \theta, z; \omega)} \nonumber \\
    \lefteqn{\hspace{40pt}= \sum_{n,m}^{n_M} v_z  (t, r, \theta, z; \omega,  k_r, m)}  \nonumber\\
    \lefteqn{\hspace{40pt}= \sum_{n, m}^{n_M} A_{n, m, \omega} J_m(k_{r;n, m} r) e^{i[k_z z + m (\theta -\theta_{m,\omega})-\omega (t-t_{n,m,\omega})]},}  \label{eq:vz_total}
\end{eqnarray}
where $A_{n,m,\omega}$  represents the real amplitude and the phase is described by $\theta_{m,\omega}$ and $t_{m,n,\omega}$. We note that $k_z$ is determined from \Eq{eq:kz} for given $\omega$, $m$, and $n$.
The inheritance is simply achieved by choosing $z$ to represent the detection layer, assuming a constant phase lag between the chromosphere and the subphotosphere in all the waves.
For reasons of practicality, we set $z=0$ to represent the detection layer, $\theta_{0,\omega} = 0$, and $t_{0,1,\omega} = 0$ for a reference without loss of generality.

\section{Data and method} \label{sec:method}

\begin{figure*}[h]
    \includegraphics[width=\hsize]{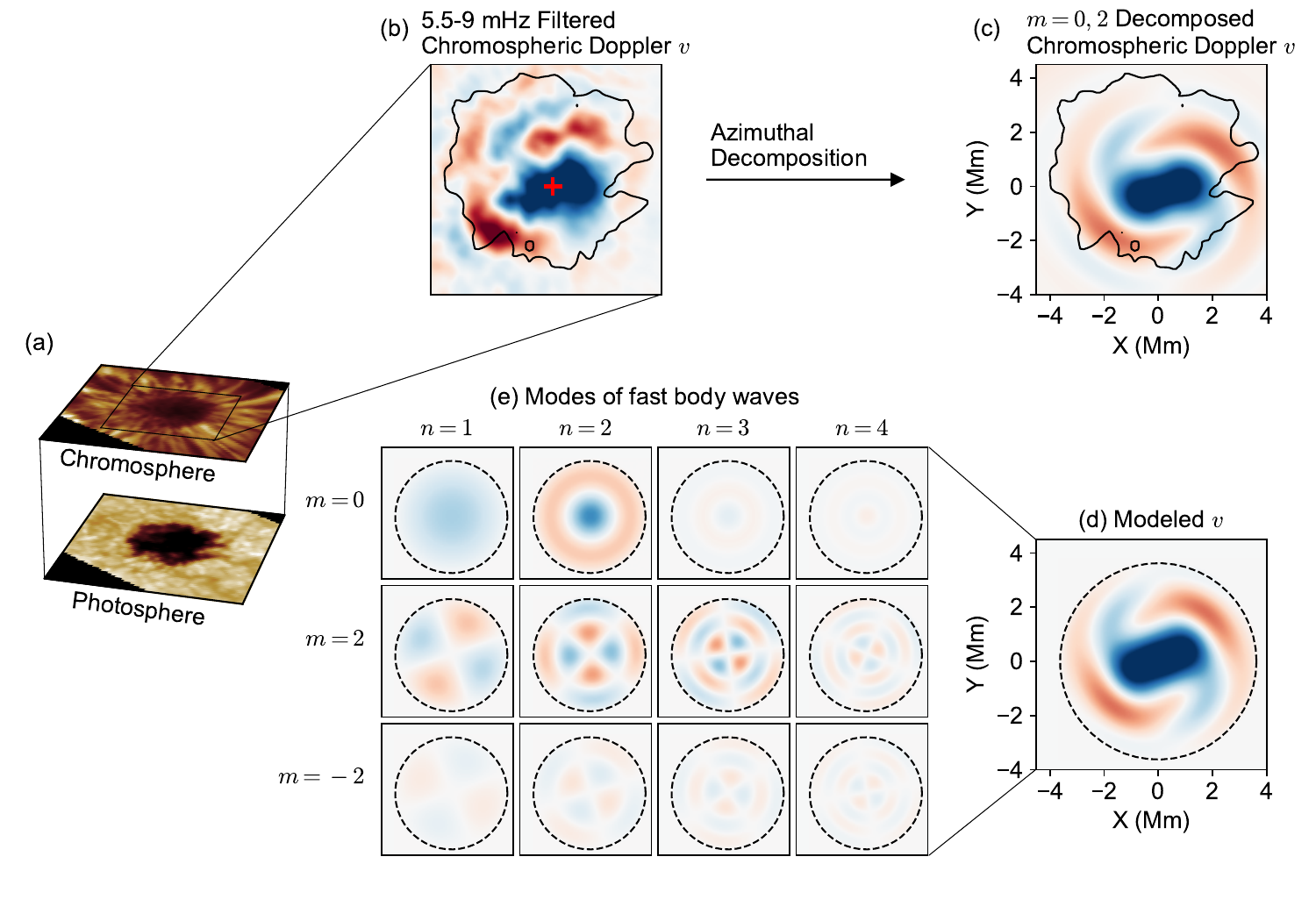}
    \caption{Detection of a pattern of umbral oscillations. 
             (a) Continuum intensity map of the GST/FISS constructed at the -3.5 \AA~ of the H$\alpha$ line center and chromospheric intensity map constructed at the line center at 17:44:07 UT on June 3, 2014.
             (b) Two-dimensional wave patterns of the chromospheric Doppler velocity map temporally filtered in $5.5-9$ mHz.
             The black contour displays the boundary of the pore, and the red cross symbol marks the center position of the SWP.
             (c) Azimuthally decomposed wave patterns for $m=0$ and $\pm 2$ modes.
             (d) Modeled LOS velocity fluctuation map constructed by the superposition of 12 modes ($m\times n$, where $m=0, 2, {-2}$ and $n=1, 2, 3, 4$).
             The dashed circle shows the boundary of the modeled flux tube.
             (e) All modes of trapped fast-body waves for the modeled LOS velocity map.
             Here $m$ represents the azimuthal mode and $n$ refers to the radial mode.
             The color limit of panels (b)-(e) is -3.5 to 3.5 km s$^{-1}$, where the positive sign signifies the redshift.
             The temporal evolutions of panels (b)-(d) are shown in Figure \ref{fig:tevo}.
            }\label{fig:overview}
\end{figure*}

\begin{figure}
    \includegraphics[width=\hsize]{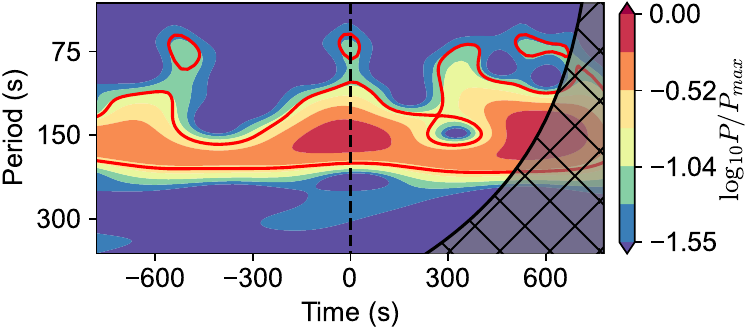}
    \caption{Wavelet power spectrum of the Doppler velocity averaged over $3\times3$ pixels at the center of the oscillation patterns.
    The black dashed line represents the time $t=0$ at 17:43:27 UT, which is the middle time of the wave packet of spiral-shaped wave patterns.
    The red contour indicates the 95\% confidence level, and the hashed region shows the cone of influence.
    Among the total observing duration (of -3200 s to 780 s), the diagram shows only the time range of -780 s to 780 s.
    }\label{fig:wavelet}
\end{figure}

We now describe how to obtain $v_z (t, r, \theta; \omega)$ in the detection layer from the observation, and how to determine the parameters $n$, $A_{n, m, \omega}$, $\theta_{m}$, and $t_{m,n}$ from $v_z (t, r, \theta; \omega)$ for each $m$ with significant power.
We applied our model to a small sunspot, a pore observed in NOAA 12078 at (-301\arcsec, 162\arcsec) on June 3, 2014, from 16:49 UT to 17:56 UT (\Fig{fig:overview}a).
The data ---taken with the Fast Imaging Solar Spectrograph (FISS, \citealp{Chae2013}) at the Goode Solar Telescope (GST)--- have been used in several studies \citep{Chae2015b,Kang2019,Chae2022}.
The time cadence of the data is 20 seconds, the spatial sampling is 0.16\arcsec, the spectral sampling is 19 m\AA, and the spectral range is -5~\AA~to 5~\AA~of H$\alpha$.
We calibrated the raw data following the reduction pipeline described in detail by \citet{Chae2013}.
We calculated the line-of-sight (LOS) Doppler velocity $v_{LOS} (t, x, y)$ at every pixel $(x,y)$ at different times using the lambdameter (bisector) method \citep[e.g.,][]{Chae2014} with a chord of 0.1 \AA.

In the present study, we focus on the two-armed SWPs that occurred around 17:44:07 UT ($t=40$ s).
The wavelet power spectrum of velocity fluctuations $v_{LOS}(t, x_c, y_c)$ at the center of these SWPs indicates that most power is concentrated in the $2-3$ minute band with a peak of around $P=160$ seconds (\Fig{fig:wavelet}).
We confined our analysis to this band, with the peak frequency $\omega= 2\pi/P = 0.04$ rad s$^{-1}$.
We extracted the velocity at this band $v_{LOS} (t, x, y; \omega)$ by applying the $5.5-9$ mHz passband filtering to the time series of LOS velocity $v_{LOS}$ at every pixel $(x,y)$ on the image plane.
As an illustration, Figure \ref{fig:overview}b shows $v_{LOS} (t, x, y; \omega)$ at $t=40$ s.

Now we describe how we extracted the azimuthal component of mode $m$ from $v_{LOS} (t, x, y; \omega)$.
We first set the polar coordinates with the origin at the center of the oscillation patterns, and obtained $v_{LOS}(t, r, \theta)$ in these coordinates using interpolation.
Second, we performed the Fourier transform of $v_{LOS}(t, r, \theta)$ over $\theta$ for every $t$ and $r$, which we then multiplied by the filter function that is set to one for the specific value of $m$ and zero for the other values.
By applying the inverse Fourier transform over $\theta$ for every $t$ and $r$, and mapping the data back to the image coordinates, we obtained the azimuthal component of velocity $v_{LOS} (t, x, y; \omega, m)$.
We found that the velocity data are satisfactorily reproduced by the summation of $m=0$ mode and $m=2$ mode only, as illustrated in \Fig{fig:overview}c.
The $m=1$ mode has not been included here because it has insignificant power \citep[see Fig. 1d in][]{Kang2019}.

To reproduce this $v_{LOS} (t, x, y; \omega, m)$, we obtained the parameters $t_{n,n,\omega}$, $A_{n,m,\omega}$, and $\theta_{m,\omega}$ from \Eq{eq:vz_total} using the following method.
First, we remapped $v_z (t, r, \theta; \omega)$ to the image coordinates, $v_z (t, x, y; \omega)$.
Second, we obtained the values of $t_{n,m,\omega}$, $A_{n,m,\omega}$, and $\theta_{m,\omega}$ for each $m$ by comparison with $v_{LOS} (t, x, y; \omega, m),$ maximizing the Pearson correlation value between the model and the observation {at a reference time of $t=40$ s}.
{The temporal evolution of the pattern is reproduced by changing time $t$ in \Eq{eq:vz_total}.}
In this process, we empirically find that $t_{n,m,\omega}$ mostly affects the overall shape of the pattern and $A_{n,m,\omega}$ reflects the width of the arms and their radial speed.
In particular, the inner shape of the patterns is highly affected by high-order $n$, and the outer part is associated with low-order $n$.
Finally, we reconstructed the patterns summing up azimuthal modes of $m=0$, $m=2,$ and $m=-2$ (\Fig{fig:overview}d).
All of the modeling parameters are shown in Table \ref{tab:params} and illustrated in \Fig{fig:overview}e.

\begin{table}
    \caption{Parameters of the model shown in Fig. \ref{fig:overview}d and Fig. \ref{fig:tevo}.}\label{tab:params}
    \centering
    \begin{tabular}{c c c c c}
        \hline\hline 
        $m$ & $n$ & $A_{m,n}$ & $\theta_m$ & $t_{m,n}$ \\
        &  & (km s$^{-1}$) & ($^\circ$) & (s) \\
        \hline
        0    & 1   & 2.02  & 0   & -3 \\
        0    & 2   & 2.94  & 0   & -25 \\
        0    & 3   & 2.13  & 0   & -55 \\
        0    & 4   & 0.51  & 0   & -69 \\
        2    & 1   & 2.42  & 30  & 30  \\
        2    & 2   & 3.13  & 30  & 2  \\
        2    & 3   & 2.42  & 30  & -41  \\
        2    & 4   & 0.72  & 30  & 68  \\
        -2   & 1   & 0.84  & 30  & 12  \\
        -2   & 2   & 1.01  & 30  & 8  \\
        -2   & 3   & 0.71  & 30  & -57  \\
        -2   & 4   & 0.75  & 30  & -79  \\
        \hline
        
    \end{tabular}
\end{table}

\section{Results} \label{sec:res}

\begin{figure*}
    \includegraphics[width=\hsize]{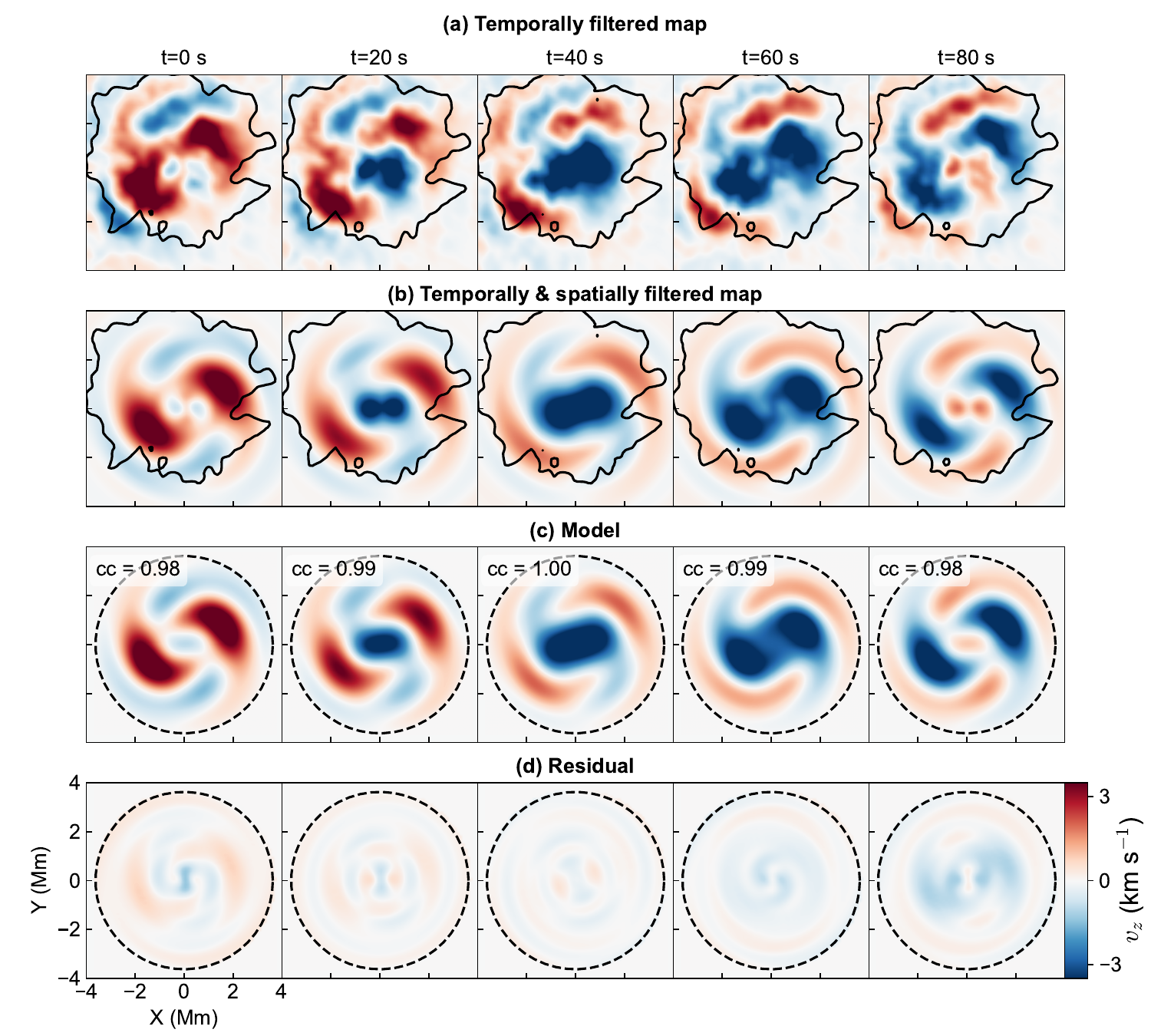}
    \caption{Temporal evolution of the spiral-shaped wave patterns.
             (a) Observed LOS velocity map temporally filtered in the frequency range of $5.5-9$ mHz.
             (b) LOS velocity map additionally filtered in the azimuthal mode of $m=0$, and 2.
             (c) Modeled velocity map constructed by the superposition of a total of 12 modes; $m=0, \pm 2$, and $n=1,2,3,4$.
             (d) Residual between the filtered map (b) and the model (c).
             The model parameters are shown in Table \ref{tab:params}.
             Columns show the temporal evolution of each map from $t=0$ to $t=80$ s.
             Here, the time $t=0$ s is equal to 17:43:27 UT, which is the middle time of the wave packet shown in Figure \ref{fig:wavelet}.
             The black solid contour indicates the boundary of the pore, and the black dashed circle represents the boundary of the flux tube of the model.
             The correlation values between (b) and (c) are shown in the top corner of each panel of (c).
             }\label{fig:tevo}
\end{figure*}

\Fig{fig:tevo} indicates that our model of subphotospheric fast-body modes can successfully reproduce the observed horizontal patterns of umbral oscillations.
Here the observed patterns have been temporally filtered in the frequency range of $5.5-9$ mHz.
The observation displays the two rotating spiral arms that appear to propagate outwards, which is successfully reproduced by the model constructed by the superposition of a total of 12 modes; $m=0, \pm 2$, and $n=1,2,3 ,4$.
As can be seen from the figure, this model is very similar in appearance to the observation, with the Pearson correlation coefficient being as large as $cc=1.00$.
The standard deviation of the residual is as small as 0.13 km s$^{-1}$ and is roughly twice the standard error of the LOS velocity of 0.07 km s$^{-1}$.
{We note that our model has the advantage that the temporal evolution is fully described with the set of model parameters computed at the reference time using \Eq{eq:vz_total}, without having to repeat the fit at individual time steps \cite[c.f.][]{Stangalini2022}.}

Table \ref{tab:params} lists the values of the model parameters.
We first consider the case of $m=2$.
In this case, the amplitude is larger than 2 km s$^{-1}$ for $n=1,2,3$ but is smaller than  1 km s$^{-1}$ for $n=4$, being as small as one-quarter of the value for $n=3$.
This means that the patterns are fairly well described by the low radial modes up to $n=3$, and including $n=4$ is more than sufficient.
In order to clarify this point, we attempted to include $n=5$.
The amplitude of the $n=5$ mode is found to be less than that of $n=4$ by a factor of 2. Furthermore, its inclusion only slightly reduces the residual error.
Therefore, we conclude that the inclusion of higher modes is physically meaningless and the observation can be successfully reproduced by a small number of modes.
The rapid decrease in the amplitude over $n$ around $n=4$ indicates that the cutoff radial mode $n_M$ may be equal to 4.
In the case of $m=0$, we may obtain a similar conclusion as in the case of $m=2$ described above.
In the case of $m=-2$, the amplitude for $n=1,2,3$ is small, not being much larger than that for $n=4$.
Nevertheless, as the amplitude of $n=4$ mode is,  in this case, comparable to the corresponding values in the cases of $m=0$ and $m=2$, and the amplitude of $n=5$ is found to be smaller than that of $n=4$, it is likely that $n_M=4$ for $m=-2$ as well.

\begin{figure}
    \includegraphics[width=\hsize]{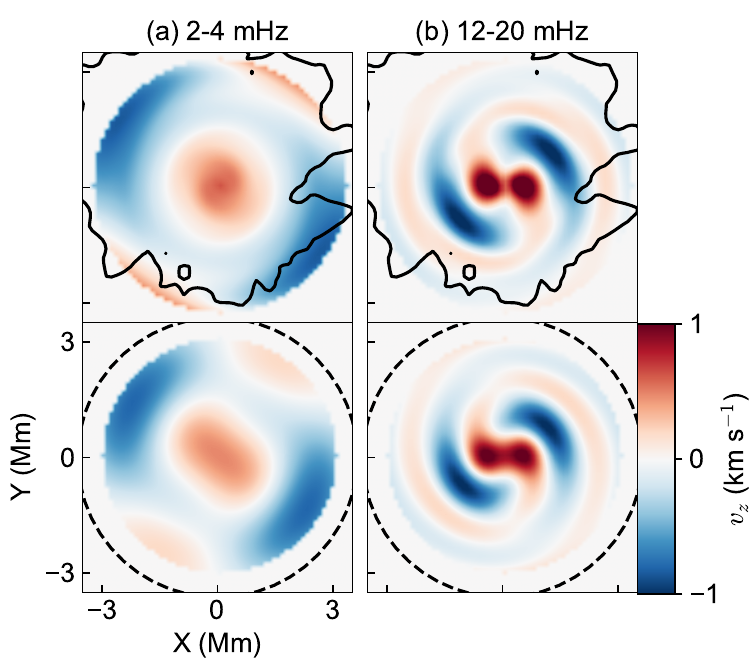}
    \caption{Oscillation patterns in low and high-frequency bands. Top panels show the observed oscillation patterns at 17:44:07 UT, which are spatially filtered in $m=0$ and 2 modes and temporally filtered in (a) $2.5-4$ mHz and (b) in $12-20$ mHz.
             Bottom panels show the model for each frequency band.
             The contour represents the boundary of the pore, and the dashed circle illustrates the boundary of the flux tube.
             All panels only show a 3 Mm radius from the center because the outer region is affected by the running penumbral waves and error.
             }\label{fig:freq}
\end{figure}

We investigated how the cutoff radial mode depends on the frequency.
To this end, we processed the observed patterns in two other bands: a lower frequency band of $2-4$ mHz ($4.2-8.3$ minute) and a higher frequency band of $12-20$ mHz ($0.8-1.4$ minute) as shown in \Fig{fig:freq}.
We find that in the lower frequency band (panel a), spiral arms are absent.
In the higher frequency band (panel b), the spiral arms not only exist but also appear to be more tightly wrapped than the $5.5-9$ mHz band pattern as shown in \Fig{fig:tevo}.
The observed patterns in these two bands can also be well reproduced by the fast-body wave models as shown in the bottom panels.
In each band, we tried different values of the maximum $n$ values for each $m$.
As a result, we found that the pattern of the $2-4$ mHz band can be reproduced satisfactorily by setting $n_M=2$, and the pattern in the $12-20$ mHz band by setting $n_M=6$.
With these values, the lower frequency pattern is well modeled with 6 modes, and the higher frequency pattern with 18 modes.
The Pearson correlation between each observed pattern and the corresponding model is found to be stronger than 0.97.

Combined with $n_M=4$ obtained above for the $5.5-9$ mHz band pattern, it is noteworthy that $n_M$ increases with frequency.
This result was empirically obtained from comparison of the models with observations and is exactly what the dispersion relation for subphotospheric fast-body modes predicts (\Fig{fig:pk}b), supporting our model of umbral oscillations inherited from subphotospheric fast-body modes.

\section{Discussion} \label{sec:dis}
In this study, we successfully reproduced the two-armed spiral-shaped wave patterns in the pore from the superposition of 12 subphotospheric fast-body modes.
We find that the patterns of umbral oscillations depend on the oscillation frequency, comparing the patterns in three different bands.
In addition, these oscillation patterns consist of the finite number of the radial mode: $n_M=4$ for the $3$ min period band, $n_M=2$ for the $5$ min band, and $n_M=6$ for the $1$ min band.
These results are in agreement with the concept of the cutoff radial mode in the model of the subphotospheric fast-body modes.

Our model is based on several assumptions for the sake of simplicity. We believe these assumptions are not detrimental.
It is assumed that the flux tube is uniform, has a circular shape, and is surrounded by an external medium of zero density. 
In Appendix \ref{appendix}, we discuss these assumptions in detail and find that they are tolerable enough to allow a reasonable interpretation of the observed patterns.
In addition, the magnetic field is assumed to be purely vertical.
Because of this assumption, we confined the modeling region to the inside of a circular pore where magnetic field inclination is almost vertical \citep{Chae2015b}.
 
{We now consider how fast waves become trapped.
In a flux tube, the fast waves can be either refracted or reflected when they reach the boundary, because the external sound speed and the internal sound speed differ from each other ($c_{s,e}>c_s$).
In particular, the waves can be fully reflected if the incident angle of the waves is larger than the critical angle ($\theta_c = \arcsin(c_s/c_{s,e})$) of the total internal reflection, similar to the light in an optical fiber.
Therefore, only the waves with an incident angle of larger than the critical angle can be fully reflected and finally form resonance modes.}

{In this regard, the cutoff in the radial mode occurs for the fast-body modes.}
In slow-body waves, the incident angles of all radial modes are large because these waves tend to be directed along magnetic fields {($k_z \gg k_r$)}.
As a consequence, all the slow waves are trapped inside the flux tube.
In contrast, in fast-body waves, with small $k_z$, {the radial directionality ($k_r/k_z$) becomes stronger for a higher $n$, because the higher $n$ has a larger value of $k_r$ by definition and has a smaller value of $k_z$ for a given period (see \Fig{fig:pk}b); this means that the incident angle becomes smaller.
As a result, the fast-body waves have a cutoff in radial modes. We note that higher radial modes with incident angles smaller than $\theta_c$ can propagate across the boundary as leaky waves.
In our case, for example, the $n=4$ mode with a period of 160 s has an incident angle of about 43$^{\circ}$, which is larger than $\theta_c\approx42^\circ$, but the incident angle of the $n=5$ mode is about 26$^{\circ}$.
Thus, only the radial mode with $n\le 4$ can be trapped in the flux tube.}

Our model is similar to the other two models of umbral oscillation patterns in several respects. 
It is like the wave propagation model in that the observed slow waves are regarded as the inheritance of the subphotospheric fast waves. It is also similar to the slow resonance model in that the oscillation patterns are described as the superposition of {several} modes.

Our model has a distinct advantage in describing the observed frequency-dependent patterns.
As shown in \Fig{fig:freq}, we find that the oscillation pattern at each frequency band is a superposition of radial modes, the number of which depends on frequency.
This frequency dependence is not at
all expected in the wave propagation model {if one single source drives the waves.}
In the wave propagation model, the horizontal pattern depends only on the depth of the source and the wave propagation speed \citep{Zhao2015,Cho2020}, and so the pattern should be the same regardless of the oscillation frequency, unless the wave propagation speed depends on frequency.
{Of course, the waves can form frequency-dependent patterns if the vertically elongated source, or more than two localized sources at different depths, generate waves of different frequencies, but this frequency dependence has not yet been reported in previous simulation works.}
The observed frequency dependence cannot be explained by the slow resonance model either.
In this latter model, an infinite number of radial modes can exist regardless of frequency \citep{Edwin1983}, which is not compatible with the finding that the pattern can be described by only a few radial modes.
In our model, in contrast, this finding is easily described by the existence of the cutoff radial mode increasing with frequency.

One might attribute the appearance of the cutoff radial mode to the dependence of the efficiency of the fast-to-slow conversion on the radial mode.
Nevertheless, this is not the case.
It is likely that the efficiency decreases with the radial mode because the conversion efficiency is known to become high when the wave vector is aligned with the magnetic field in order to ensure $k_z/k_r \gg 1$ \citep{Schunker2006}.
However, this expected decrease is gradual, unlike the observed sharp drop in amplitude (see Table \ref{tab:params}), which is naturally interpreted as the occurrence of the cutoff.

\begin{figure}
    \includegraphics[width=\hsize]{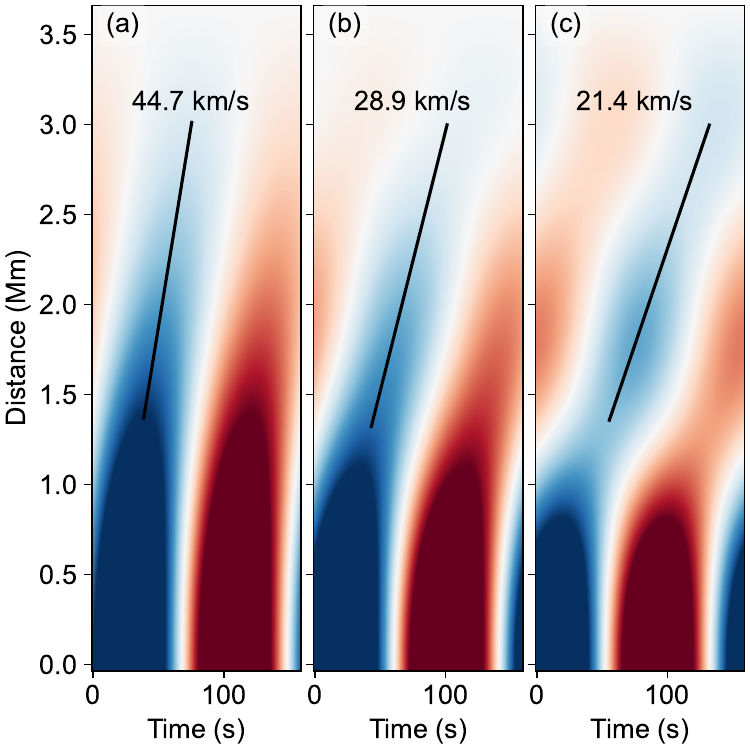}
    \caption{Time--distance map of modeled $m=0$ mode waves for three different cases.
             (a) TD map with the amplitudes of $A_{0,1} = 2.25$, $A_{0,2} = 2.0,$ and $A_{0,3} = 0.5$ in units of km s$^{-1}$.
             (b) TD map with the amplitudes of $A_{0,1} = 1.5$, $A_{0,2} = 2.25$ and $A_{0,3} = 1.5$ in units of km s$^{-1}$.
             (c) TD map with the amplitudes of $A_{0,1} = 0.5$, $A_{0,2} = 2.0$ and $A_{0,3} = 2.25$ in units of km s$^{-1}$.
             All of them have the phase differences of $t_{0,1} = 0$, $t_{0,2} = -20$ and $t_{0,3} = -40$ in seconds.
             The solid lines represent the gradients of the ridges, and their values are shown in each panel.
             }\label{fs:tdmap}
\end{figure}

Our model requires the trapping of fast waves in the subphotosphere.
We note that the fast-body modes can exist at regions satisfying the condition of \Eq{eq:fastcon}.
This condition is not satisfied in the photosphere, or above it, where $v_A>c_{s,e}>c_s$.
Nor is it satisfied in the deep region where  $c_s \sim c_{s,e}$.
Therefore, it is likely that the trapping of fast waves is occurring in the shallow region in the subphotopshere.
Indeed, we estimated the trapping region by calculating the sound speed and the Alfv\'en speed in the subphotosphere using the Maltby model \citep{Maltby1986}, and found that the fast-body modes of the observed two-armed SWPs may be formed in a shallow region from -300 km to -50 km below the equipartition layer.
In this region, plasma $\beta$ is found to range from 3 to 10.
We think that this region is for horizontal trapping only, and is not for vertical trapping. In the vertical direction, fast waves might come from below or from above.

Our model of fast-body modes has a couple of noticeable characteristics. First,
the apparent radial propagation of the pattern is reproduced by an appropriate combination of amplitudes in the radial modes.
For example, Time--distance (TD) maps in Figure \ref{fs:tdmap} clearly show that the speed of radial propagation increases with the ratios of the $n=1$ mode amplitude to that of the other modes.
In this regard, our model is in contrast with the wave propagation model, where the speed of the radial motion depends on the depth of the localized source \citep{Zhao2015,Felipe2017,Kitiashvili2019,Cho2020}.
Second, fast-body modes, in principle, can be excited by arbitrary sources, either internally or externally.
{Among these two possible sources, we conjecture that} the predominant source may be the external $f$ and $p$ modes in the subphotosphere, {because $f$ and $p$ modes ubiquitously occur in the quiet Sun region regardless of the existence of a sunspot, and the $p$-mode absorption is an effective mechanism to absorb incoming p-mode waves \citep[e.g.,][]{Braun1987,Cally1997}.}
In addition, irrespective of the excitation source, the fast-body modes, once established, display the pattern of umbral oscillations apparently propagating out of the center of the cross-section of the flux tube.
In the wave propagation model, in contrast, the source is assumed to be a localized disturbance in the subphotosphere, which is related to the center of the oscillation patterns observed in the chromosphere \citep{Cho2020, Kang2024}.

Our model has difficulty in explaining the observed change of the center of the pattern.
From the observation of a sunspot with the light bridge, \citet{Cho2021} reported that the center of the wave pattern was located at different positions inside the same umbra at different times.
It is difficult to explain this observation with our model as far as we currently understand, unless the shape of the sunspot significantly changed during the observation.
This is a challenging problem for our model.
To obtain a solution, further numerical and observational studies are necessary.

\begin{acknowledgements}
{We appreciate the referee's positive evaluation and the constructive comments.}
This research was supported by the National Research Foundation of Korea (RS-2023-00273679).
J. Chae was supported by the National Research Foundation of Korea (RS-2023-00208117).
E.-K. Lim was supported by the Korea Astronomy and Space Science Institute under R\&D program {of the Korean government (MSIT; No. 2024-1-850-02, 2024-1-810-03).}
We gratefully acknowledge the use of data from the Goode Solar Telescope (GST) of the Big Bear Solar Observatory (BBSO). {BBSO operation is supported by US NSF AGS-2309939 and AGS-1821294 grants} and New Jersey Institute of Technology.
GST operation is partly supported by the Korea Astronomy and Space Science Institute and the Seoul National University.
\end{acknowledgements}

\bibliographystyle{aa}
\bibliography{kang}

\begin{appendix}

\section{Dependence of the model on the assumptions} \label{appendix}
\begin{figure}
    \includegraphics[width=\hsize]{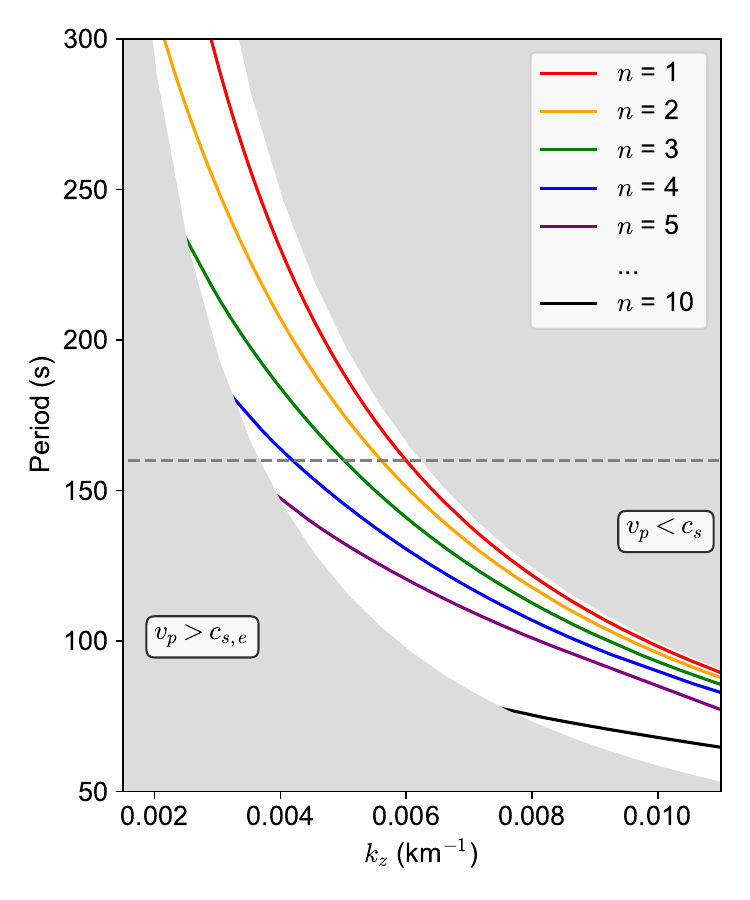}
    \caption{Numerical solution of \Eq{eq:dispersion} with the same parameters as Figure \ref{fig:pk}b but without the assumption of zero-density limit.
                As the fast-wave solution is confined within the range of $c_{s,e}>v_p>c_s$, curves are confined to the white area.
                }\label{fig:app}
\end{figure}
Our model, for simplicity, assumes the zero-density limit, circular shape, and uniformity inside the flux tube.
Here we discuss how each of these assumptions affects the model.

\subsection{Zero-density limit}
As a boundary condition, our model assumes that mass density is zero outside the flux tube.
In this zero-density limit, the node is always located at the boundary of the flux tube, which leads to the very simple boundary condition of \Eq{eq:simple}.
However, in the general case, the external density is not zero and, hence, the node does not occur exactly at the boundary.
To examine the difference, we solve the full boundary condition in \Eq{eq:dispersion} and present the solution in \Fig{fig:app}.
As a result, we find that the node is located slightly inside the boundary.
The mode has slightly larger values of $k_r$, $k_z$ for a given period than in the case of zero-density limit (see \Fig{fig:app}).
An important point here is that the differences are sufficiently small, as can be seen from the similarity of the curves shown in \Fig{fig:app} to those in \Fig{fig:pk}b.
This result suggests that the assumption of a zero-density limit is sufficiently justified to allow interpretation of the oscillation patterns, unless precise modeling is required.

\subsection{Circular shape}
The geometry of the flux tube can change the shape of the pattern \citep{Albidah2022}.
If the cross-section of the flux tube is not circular, the center of the oscillation pattern is not located at the center of the flux tube.
In our study, the effect of noncircular shape does not seem to be strong because the pore has a relatively circular shape.
In our model, we identified the center of the two-armed SWPs at the center of the flux tube, which is found to be 1\arcsec\  from the center of the pore determined by the ellipse fitting of the morphology.
We think the deviation of this amount is tolerable, and so the assumption of the circular shape is sufficiently justified for our purposes.
We also note that the geometry of the flux tube does not affect the presence of the cutoff radial mode, which is the most important characteristic of our model here.

\subsection{Uniformity}
For simplicity, our model assumes that all the physical parameters are uniform both horizontally and vertically in the subphotosphere.
The assumption of uniformity holds better in the horizontal direction than in the vertical direction.
The vertical uniformity is hindered by the gravitational stratification of pressure and density.
Nevertheless, it may not be detrimental in our model because the vertical extent of the subphotosphere ($\sim 250$ km) is not much larger than the local pressure scale height ($\sim 200$ km).

\subsection{Trapping}
The existence of the modes implicitly assumes that the waves are trapped somehow inside the flux tube.
The trapping requires not only horizontal confinement but also vertical confinement.
It is obvious that horizontal confinement occurs at the lateral boundaries.
This leads us to question where the vertical confinement might occur.
Our model of fast-body modes implicitly assumes that the vertical confinement occurs outside the subphotosphere: both above and below it.
In the region above the photosphere, density rapidly decreases with height and the Alfv\'en speed rapidly increases with height, which causes the fast waves to be reflected downward.
In the deep interior below the subphotosphere of our interest, the temperature rapidly increases with depth, and the propagation speed of the fast waves rapidly increases with depth, which causes the waves to be reflected upward.

\end{appendix}

\end{document}